\documentstyle[epsfig]{aipproc}
\begin{document}
\title{Lepton Flavour Violation Experiments -- Some Recent Developments}
\author{Klaus P. Jungmann}
\address{Physikalisches Institut, Universit\"at Heidelberg\\
Philosophenweg 12, D-69129 Heidelberg, Germany}
\maketitle
\begin{abstract}
Dedicated experiments searching for lepton flavour violation
can be performed very sensitively using  K-decays and  $\mu$-decays
as well as neutrinoless double $\beta$-decay and
muonium to antimuonium conversion. Although there is no confirmed signal
reported yet, stringent limits for parameters in speculative extensions to
the standard model can be set. Some models could recently be ruled out.
\end{abstract}

\section{Introduction}

All confirmed experimental data acquired to date indicate
the conservation of lepton numbers.
This fact can be described by several different empirical laws
\cite{Zel_52,Pon_59,Cab_60,Kon_53,Fei_61}, some of which follow additive
and some obey multiplicative, parity-like, schemes.
Experiments have given no indication yet for favouring any of them.
The standard model states for every lepton flavour a separate additively conserved
quantum number.
However, such lepton numbers have no status,
unless their conservation can be associated with a local gauge invariance
\cite{Halp_93}.
Mixings between different generations are well known in the quark sector and
the Cabbibo-Kobayashi-Maskawa matrix \cite{Kob_73} relates the weak
quark eigenstates with their mass eigenstates. A familiar example are the
${\rm K}^0$-$\overline{{\rm K}^0}$ oscillations.
At present we are left  puzzled why leptons do not show any similar mixing.
Recent
experimental hints for neutrino
oscillations, which have a potential for changing this situation,
are not covered here (see e.g. \cite{Stone_98}).

Many extensions to the standard model have been proposed and are presently
discussed which try to explain further
some of its not well understood features like
e.g. parity violation in weak interaction or particle mass spectra.
They are  put by hand into this remarkable theoretical framework
which appears to serve as an extremely robust description of all
confirmed particle physics.
Lepton flavor violation (LFV) appears naturally in such models
which include Left-Right-Symmetry, Supersymmetry, Technicolor, Grand Unification,
String Theories, Compositeness, and many others. They continue to stimulate
experimental searches in a large range of energies.
\begin{figure}[thb]
\unitlength 1.0 cm
 \begin{picture}(11.0,7.0)
  \centering
   \epsfig{figure=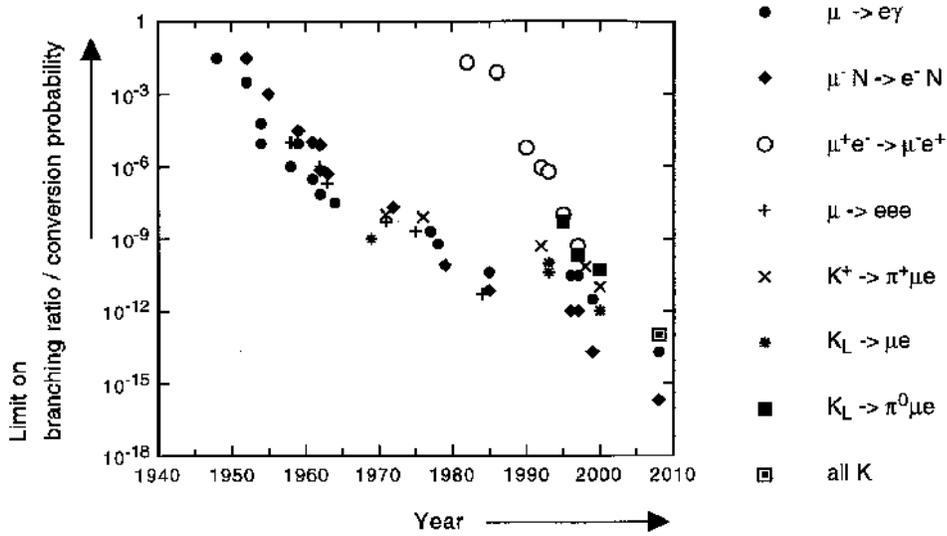,height=7.0cm,angle=270}
 \end{picture}
 \centering\caption[]
        {\protect{\label{history}}
        Dedicated searches for lepton number violating processes
         involving muons ($\mu$) and kaons (${\rm K}$).
         Recent ${\rm K}$ experiments and $\mu^+e^-$ -- $\mu^-e^+$
         conversion exhibit the most significant gains in sensitivity. Points
         in 1998 and beyond are projections of possibilities by the respective
         experimenters.
         }
\end{figure}
 With some low energy experiments
new physics can be probed at mass scales far beyond the reach of
present accelerators or such planned  for the future.

Highest sensitivity has generally been reached in dedicated search experiments
particularly on Kaons (${\rm K}$) and muons ($\mu$) (Table \ref{lnv_limits}),
where also a high discovery potential for new physics exists \cite{Mohapatra_93},
as well as in non accelerator experiments searching for neutrinoless
double $\beta$-decay. The  decays of heavier objects
created in high energy collisions, however, can be observed less accurately.
The progress in the ${\rm K}$ and $\mu$
(see sec. \ref{kaons} and \ref{muons})
field is indicated in Fig.\ref{history}
which shows more than 10 decades of improvement since the first experiments
in the late 1940's. The highest recent gain in sensitivity
is for muonium (M=$\mu^+e^-$) to antimuonium ($\overline{\rm M}$=$\mu^-e^+$)
conversion due to a new, yet unused signature (see sec. \ref{mmbar}).
%
%
\begin{table}[hbt]
 \caption[]{\protect{\label{lnv_limits}} Recently obtained upper limits on
lepton number violating processes (90\% C.L.).}
\begin{tabular}{|clcc||clcc|}
 \hline
decay && limit && decay && limit& \\
 \hline
Z$^0 $&$\rightarrow \mu e   $ & $2.5  \cdot 10^{-6}$ &\cite{adr_93}
&K$_L$&$\rightarrow \mu e   $ & $2  \cdot 10^{-11}$ &\cite{Molzon_98}\\
Z$^0 $&$\rightarrow \tau e  $ & $7.3\cdot 10^{-6}$ &\cite{adr_93}
&K$_L$&$\rightarrow \pi^0 \mu e $& $3.2  \cdot 10^{-9}$ &\cite{ans_97}\\
Z$^0 $&$\rightarrow \tau \mu $& $1.0\cdot 10^{-5}$ &\cite{adr_93}
&K$^+$&$\rightarrow \pi^+ \mu e $& $4  \cdot 10^{-11}$ &\cite{Zeller_98}\\
D$^0 $&$\rightarrow \mu e  $  & $1.9  \cdot 10^{-5}$ &\cite{Frey_96}
&$\mu^+$&$\rightarrow e^+ \nu_{\mu} \overline{\nu}_e$ & $2.5  \cdot 10^{-3}$ &\cite{Eitel_95}\\
D$^0 $&$\rightarrow \pi^0 \mu e $ & $8.6 \cdot 10^{-5}$ &\cite{Frey_96}
&$\mu$&$\rightarrow eee     $ & $1  \cdot 10^{-12}$ &\cite{Bertl_85}\\
D$^0 $&$\rightarrow \Phi  \mu e $ & $3.4 \cdot 10^{-5} $&\cite{Frey_96}
&$\mu$&$\rightarrow e \gamma  $   & $3.8 \cdot 10^{-11}$ &\cite{Cooper_97}\\
B$^0 $&$\rightarrow \mu  e  $  & $5.9 \cdot 10^{-6} $&\cite{Ammar_94}
&$\mu^-$Ti&$\rightarrow e^- $Ti      & $6.1 \cdot 10^{-13}$ &\cite{Eggli_98}\\
B$^0 $&$\rightarrow \tau e   $ & $5.3 \cdot 10^{-4} $&\cite{Ammar_94}
&$\mu^-$Ti&$\rightarrow e^+ $Ca      & $1.7 \cdot 10^{-12}$ &\cite{Kaulard_98}\\
B$^0 $&$\rightarrow \tau \mu $ & $8.3 \cdot 10^{-4} $&\cite{Ammar_94}
&$\mu^+ e^-$&$\rightarrow \mu^- e^+$ & $ G_{\rm M \overline{M}}<3 \cdot 10^{-3} G_F$
&\cite{Vmeyer_97}\\
B$^0 $&$\rightarrow$ K $ \mu e  $& $1.8 \cdot 10^{-5} $&\cite{Ammar_94}
&$^{76}$Ge &$\rightarrow ^{76}$Se~~ $e^-e^-$ & $ T_{1/2} > 1.2 \cdot 10^{25} y$ &\cite{Klap_98}\\
$\tau$&$\rightarrow e \gamma  $   & $2.7 \cdot 10^{-6} $&\cite{Edwards_97}
&&&$m_{\nu_e}(Maj.) < 0.45 eV$ &\cite{Klap_98}\\
$\tau$&$\rightarrow \mu \gamma  $   & $3.0 \cdot 10^{-6} $&\cite{Edwards_97}
&&&&\\
\hline
\end{tabular}
\end{table}

\section{Neutrinoless double $\beta$-decay}
\begin{figure}[bht]
\unitlength 1.0 cm
 \begin{picture}(11.0, 6.0)
  \centering
   \epsfig{figure=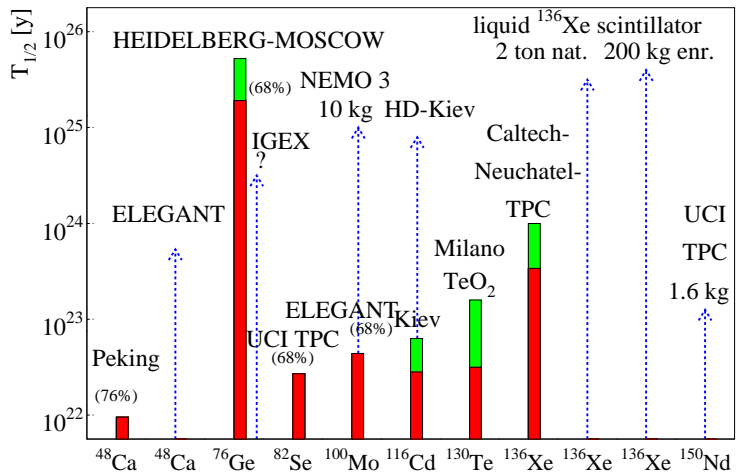,height=6.2cm}
 \end{picture}
 \hspace{0.2 cm}
 \centering\caption[]{\protect{\label{double_beta}}
         Experiments searching for neutrinoless double
        $\beta$-decay. The most sensitive ones use enriched $^{68}$Ge detectors.
        The dark areas represent the current status and the lighter color
        indicates near future possibilities. The dashed arrows are long term future
        plans. Among the most ambiguous projects ranks a 1 ton Ge detector
        which could be used to gain two orders of magnitude in sensitivity
        (from ref.\protect{\cite{Klap_98}}).}
\end{figure}

A $\beta$-decay of a nucleus involving two electrons and no neutrinos would
violate electronic lepton flavour by two units. It has been suggested in
many models, particularly, such involving neutrinos of Majorana type.
It is  being  searched for in many experiments (see Table \ref{double_beta})
using $^{48}$Ca, $^{76}$Ge, $^{82}$Se , $^{100}$Mo, $^{116}$Cd, $^{130}$Te
and $^{136}$Xe. Among those the Heidelberg-Moscow
Germanium experiment provides the
most stringent half life limit of $T_{1/2} \geq 1.2\cdot 10^{25}$ y
\cite{Klap_98}.
It uses most advantageously
isotopically enriched material with 86\% $^{76}$Ge as a semiconductor
detector to watch its own nuclei decay.
It is situated in  a clean and carefully against background radiation shielded
environment in the Gran Sasso underground laboratory. The measures include
purging with purified nitrogen as well as using
copper material in the cooling system in the vicinity of the actual detector
which was selected for low intrinsic radiation.
Remaining background counts were further suppressed by pulse shape
analysis. The result achieved in 31.8 kg years with the 11.5 kg detector
can be used to impose an upper limit on the electron neutrino Majorana
mass of 0.45 eV, which is well below the electron neutrino mass limit
of 3.9 eV established in model independent general direct searches using
tritium decay \cite{Lobashev_97}.

With 1 ton enriched $^{76}$Ge distributed in 288 individual detectors,
as suggested by the GENIUS proposal, one could expect in 10 years running
time a limit of $T_{1/2} > 6\cdot 10^{28}$ y corresponding
to a Majorana neutrino mass limit of below 6 meV/c$^2$ \cite{Klap_98}.
It is a particularly nice feature of most experiments searching for
neutrinoless double $\beta$-decay that they can also contribute to sensitive
searches for cold dark matter, particularly weakly interacting massive
particles (WIMPS) in mass regions above $\approx$ 20 GeV/c$^2$.

\section{Experiments at electron-positron colliders - Z$^0$ and W$^\pm$ bosons
and $\tau$ leptons}

The general purpose detectors installed at the large high energy
electron-positron colliders provide  the possibility
heavy elementary particles and gauge bosons
like the $\tau$ lepton or the $W^{\pm}$ and $Z^0$
bosons   to be observed for rare decays and particularly for lepton number violating
effects. Their high mass offers  for each particle a large number of different
possible purely leptonic and semileptonic decay channels.
Z$^0$ bosons were produced in large quantities at the LEP storage ring of CERN
and the Stanford Linear Collider. With the LEP200 upgrade a significant number
of W$^\pm$ bosons became available. For $\tau$'s the CLEO detector
at the Cornell CESR facility provided a significant amount of the available data
particularly on neutrinoless $\tau$ decays \cite{Edwards_97,Bliss_98}
as well as on ${\rm B}^0$ and ${\rm D}^0$ decays \cite{Frey_96,Ammar_94}.

The sensitivity of all analyses
for lepton number violating (LNV)
decays have a principle limit set by statistics. For a particular
decay channel further restrictions arise from finite acceptances
for the final state particles
which explain
the course differences in the upper bounds  reported for the different
channels although starting from the same initial state (Table \ref{lnv_limits}).
The limits on branching ratios are in general much higher than the ones obtained in
dedicated experiments on ${\rm K}$'s and $\mu$'s. For $\tau$'s one expects
in the near future a sensitivity not better than $10^{-7}$ for
any decay mode.

However, such bounds are of great value
for discriminating theoretical models where mass scaling runs with a
high power of the mass ratios. In the framework of superstring models,
for example,
the decay $l \rightarrow e \gamma$, where $l$ stands for $\mu$ or $\tau$,
scales with the fifth power of the lepton mass $m_{l}$. In this particular case
the upper limit of
$2.7\cdot 10^{-6}$ for $\tau \rightarrow e \gamma$ can compete with
the present $3.8 \cdot 10^{-11}$
limit on $\mu \rightarrow e \gamma$ due to the $1.3 \cdot 10^{6}$ enhancement
factor from the mass ratio $m_{\tau}/m_{\mu}\approx16.8$.
 However in general the mass scaling is expected to be less dramatic.

\section{Experiments on Kaons}

\protect{\label{kaons}}
With the availability of intense Kaon sources at the
Fermi National Accelerator Laboratory (FNAL)
and
the Brookhaven National Laboratory (BNL) and with novel
experimental techniques
developed to cope with high data rates the search for LFV K-decays
has gained a lot of interest. Here the experiments
BNL-871 searching for  ${\rm K}^+ \rightarrow \pi^+ e \mu$,
BNL-865 searching for ${\rm K}_L \rightarrow \mu e$
and the Fermilab KTeV effort FNAL-799II investigating
${\rm K}_L \rightarrow \pi^0 \mu e $ promise significant improvements
(see Table \ref{K_decays}), where the LFV decays are searched along
with measurements on very rare K decay channels.
Among the new physics that could be
revealed are new heavy gauge bosons with masses up to order
50-200 TeV/c$^2$, far beyond the reach of even any planned accelerator
\cite{Kettell_98}. At the projected Japanese Hadron Facility (JHF)
one could expect significant improvements
beyond the present status.

%
\begin{table}[bht]
\centering
 \caption[]{\protect{\label{K_decays}} Three presently ongoing searches for
lepton flavour violating K decays.}
\begin{tabular}{|l||c|ccc|}
\hline
&\hspace*{3mm}past limit \hspace*{3mm}& \hspace*{3mm}present limit & \hspace*{3mm}anticipated limit of & future \\
&           &  (1998)       &  ongoing experiment  & possibility \\
 \hline
\hspace*{3mm}${\rm K}^+ \rightarrow \pi^+ e \mu$
& $2\cdot 10^{-10}$ &$ 4\cdot 10^{-11}$&$\approx 3\cdot 10^{-12}$ & $ 10^{-13}$\\
& BNL-777 &&BNL-865&\\
\hspace*{3mm}${\rm K}_L \rightarrow \mu e$
& $3\cdot 10^{-11}$ &$ 5\cdot 10^{-12}$&$\approx 8\cdot 10^{-13}$ & $ 10^{-13}$\\
& BNL-791 &&BNL-871&\\
\hspace*{3mm}${\rm K}_L \rightarrow \pi^0 \mu e $
& $3.2\cdot 10^{-9}$ &          &$\approx 1\cdot 10^{-11}$ & $ 10^{-13}$\\
& FNAL-799 &&FNAL799II&\\
\hline
  \end{tabular}
\end{table}

\section{Experiments involving Muons}

\protect{\label{muons}}

The decay $\mu \rightarrow e\gamma$ was the first being searched
for shortly after the muon's nature as a heavy electron-like particle became
apparent. Searches for rare and forbidden muon decays
have been among the most precise
experiments in physics since and have always been of special interest
in the context of unified gauge theories, as they can provide accurate
tests of speculative models and because of the achievable experimental
precision they may be able to discriminate between such
\cite{Vergados_94}.  Recently forbidden muon decays have attracted
attention, when their possible sensitivity to effects
arising in minimal supersymmetry (SUSY)
were discussed in theoretical studies
\cite{Barbieri_95}.
It was pointed out that
for values of $tan{\beta}$ ( the ratio of the vacuum expectation values of the
two Higgs fields involved) which exceed about 3, the branching ratio
should be above $\approx 10^{-14}$ for a decay $\mu \rightarrow e\gamma$ and
above $\approx 10^{-16}$ for $\mu \rightarrow e$ conversion on a Ti nucleus,
almost independent
of all other parameters in the model. This has stimulated a letter of intent
to the Paul Scherrer Institute (PSI), Switzerland,
and a proposal to BNL to search for the respective processes.

In the field of searching for SUSY effects in low energy experiments
rare decay experiments
are in some competition with the just started new
precision measurement \cite{Miller_97}
of the muon magnetic anomaly $a_{\mu}$ where the contribution from
SUSY is of order
$a_{\mu}(SUSY)=140\cdot 10^{-11} tan{\beta} * (100 GeV/\tilde{m})^2$
with $\tilde{m}$ the mass of the lightest SUSY particle (see \cite{Nath_95}).
The measurement goal is $\Delta a_{\mu}(exp)=40\cdot 10^{-11}$ and should
be reached around the year 2001.

\subsection{$\mu \rightarrow e\gamma$ decay}

The signature of a $\mu \rightarrow e\gamma$ event is a 52.8 MeV positron emitted
back to back with a 52.8 MeV photon. The MEGA experiment at the late Los Alamos
Meson Physics Facility (LAMPF) consisted of a magnetic spectrometer to observe
the charged final state particle and three pair spectrometers for detecting
the photon through its $e^+$$e^-$ pair  creation in lead converters.
Random coincidences at high rates are reported as major background. Data taking
is completed and 16\% of the data could be analyzed leading to an upper
limit on the branching ratio of $3.8 \cdot 10^{-11}$ \cite{Cooper_97} which
slightly improves the value of $4.9 \cdot 10^{-11}$
established in a crystal box detector also at LAMPF
\cite{Bolton_88}.

At PSI new efforts are being discussed to reach a sensitivity of
about $5 \cdot 10^{-14}$
for this decay mode within the next couple of years.
The suggested instruments include solutions like
a large solid angle magnetic spectrometer for the $e^+$ surrounded by a crystal
calorimeter for the $\gamma$, or liquid Xe calorimeters for the   $\gamma$
and others \cite{vds_98}.

It should be noted that the tightest bounds on bileptonic gauge bosons,
which are common to many speculative standard model extensions, come from
$\mu \rightarrow e\gamma$, if flavour democracy is assumed \cite{Cuyp_96}.

\subsection{$\mu \rightarrow e $ conversion}

Many constraints for speculative models arise from the present
experimental bound on the conversion process $\mu + Z \rightarrow e + Z$
(Table \ref{lnv_limits}),
which is the tightest for all studied LNV decays.
The variety of theoretically possible processes that can be tested
includes, e.g.
supersymmetric loop graphs, heavy neutrinos, leptoquarks, compositeness, Higgs bosons
and heavy $Z'$  bosons with anomalous couplings. Generally it is more
sensitive to new Physics  than $\mu \rightarrow e \gamma$ in a wide class
of models where the process is generated at the one loop level \cite{Raidal_97}.

The process needs to involve a nucleus to assure elementary conservation laws.
If the nucleus is left in its ground state, a conversion event is
signaled through
the release of a 105 MeV electron, which is uniquely distinguishable from
normal muon decay electrons ranging up to 53 MeV. Among the physically
relevant intrinsic background processes is $\mu$ decay in the
atomic orbit after
a muonic atom has been formed,
which can release much higher energetic electrons,
and radiative muon capture, where the photokinematic end point can be close to
the signal electron energy.

The ongoing SINDRUM II experiment uses the worldwide brightest continuous
muon channel $\pi$ E5 at PSI. Their new results limit tyhe branching ratios
$\mu^- {\rm Ti}\rightarrow e^+ {\rm Ca}^{gs}$ to  below $ 1.7\cdot 10^{-12}$
for the {\rm Ca} nucleus in the ground state \cite{Kaulard_98},
$\mu^- {\rm Ti}\rightarrow e^+ {\rm Ca}^{GDR} $ to below $ 3.6\cdot 10^{-11}$ leaving Ca with
giant dipole resonance excitation \cite{Kaulard_98}, and
$\mu^- {\rm Ti}\rightarrow e^- {\rm Ti}$ to below $ 6.1\cdot 10^{-13}$ for Ti in the
ground state \cite{Eggli_98}.
For the ground state processes the nucleons interact coherently
which enhances the possible effect.
In order to  boost accuracy in the near future the SINDRUM II collaboration
wants to take advantage in the gain of muon flux through
a $\pi -\mu$ converter, a novel superconducting device in the beam line
which collects $\pi$'s and releases only $\mu$'s
with very low $\pi$ contamination. The latter point is essential
as $\pi$'s are a source of potential background due to nuclear reactions. The
projections of the collaboration
for the achievable limit in the coherent $\mu^- {\rm Ti}\rightarrow e^- {\rm Ti}$
case are in the $10^{-14}$ region.

\begin{figure}[t]
\unitlength 1.0 cm
 \begin{picture}(11.0,7.0)
  \centering
   \epsfig{figure=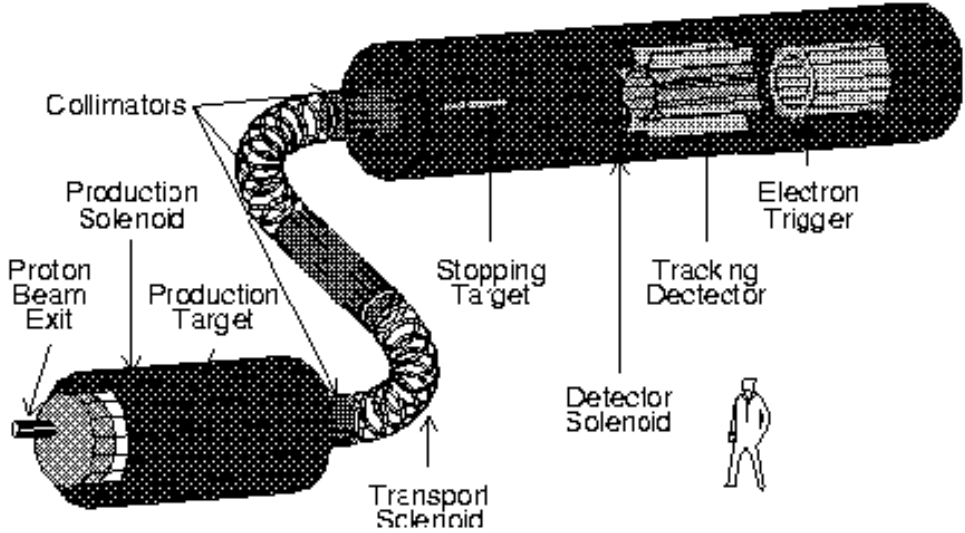,height=7.0cm}
 \end{picture}
 \centering\caption[]{\protect{\label{MECO}}The MECO experiment planned at BNL
 (see \cite{Molzon_97}).}
\end{figure}

The new Muon Electron Conversion (MECO) experiment proposed at BNL \cite{Molzon_97}
(see Fig.\ref{MECO}) is very close
in its design to a proposal by Lobashev and collaborators for
the Moscow Meson Factory.
The setup consists of a target station for $\pi$/$\mu$ production which uses
a proton beam from the AGS accelerator,
an S-shaped  transport and purification section
and a detector the basic idea of which
is to let electrons from normal muon decay pass without
being seen and to observe
only the 105 MeV signal electrons.
The goal is  the $10^{-16}$ level
in sensitivity, which will stringetly test supersymmetric
models; there is an anticipated ultimate capability for $10^{-18}$
\cite{Kirk_98}.

\subsection{$\mu^+ e^- \rightarrow \mu^- e^+ $ conversion}

\protect{\label{mmbar}}
The hydrogen-like muonium atom consists of two leptons from
different generations.
The close confinement of the bound state offers excellent opportunities to
explore precisely fundamental electron-muon interactions
\cite{Hughes_90,Jungmann_94}.
Since the effect of all known fundamental forces in this system are calculable
very well mainly in the framework of quantum electrodynamics (QED), it
renders the possibility to search sensitively for yet unknown
interactions between both particles.
\begin{figure}[bht]
\unitlength 1.0 cm
 \begin{picture}(11.0,3.8)
  \centering
   \hspace{0.9 cm}
   \epsfig{figure=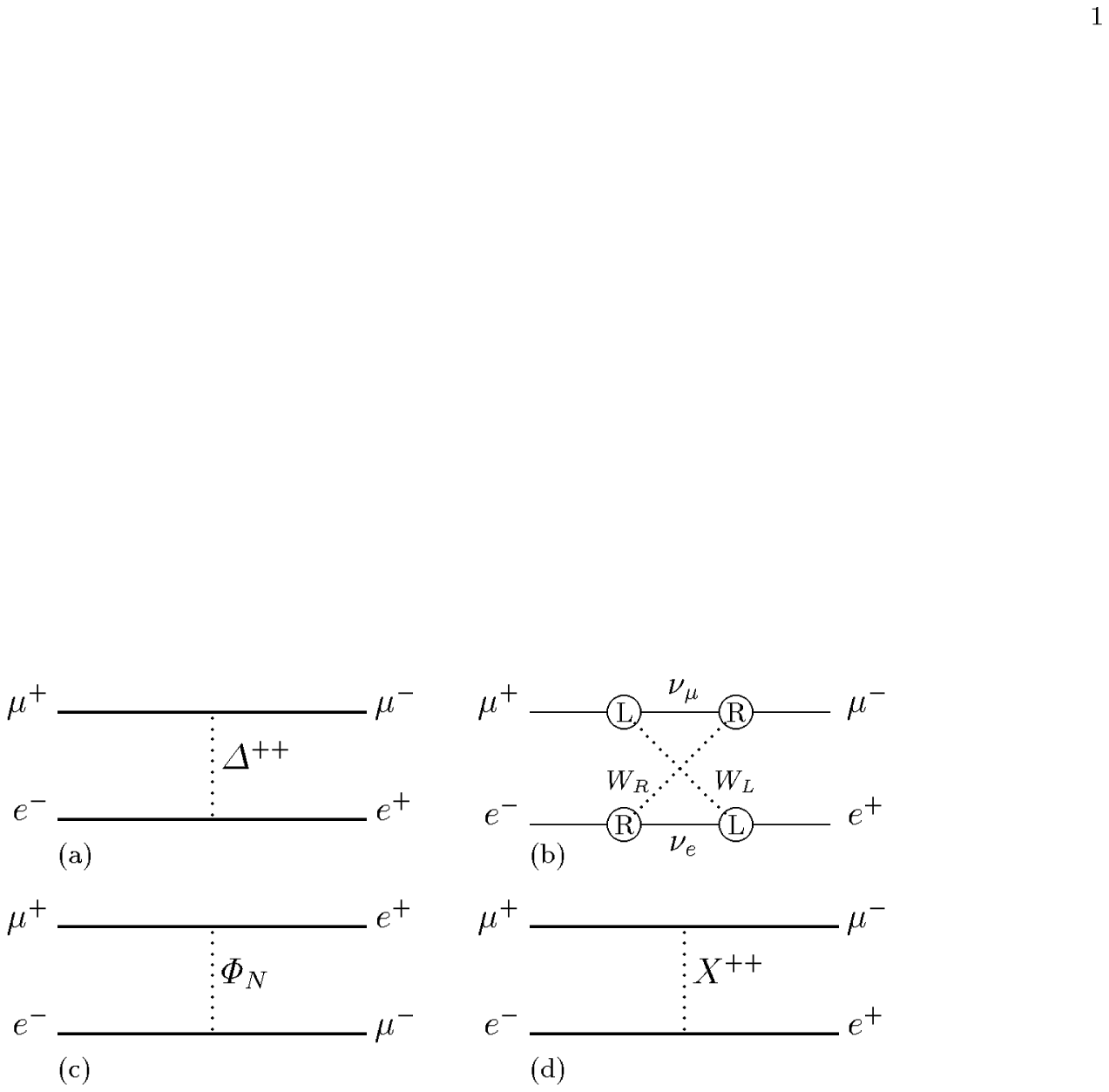,height=10cm}
 \end{picture}
 \centering\caption[]
        {\protect{\label{theo_mmb}}
        Muonium-antimuonium conversion in
        theories beyond the standard model. The interaction
        could be mediated by
        (a) a doubly charged Higgs boson $\Delta^{++}$
        \protect{\cite{Halprin_82,Herczeg_92}},
        (b) heavy Majorana neutrinos \protect{\cite{Halprin_82}},
        (c) a neutral scalar $\Phi_N$ \protect{\cite{Hou_96}}, e.g.
        a supersymmetric $\tau$-sneutrino $\tilde{\nu}_{\tau}$
        \protect{\cite{Halp_93,moha92}}, or
        (d) a dileptonic gauge boson $X^{++}$ \protect{\cite{Sasaki_94}}.
        }
\end{figure}

An M-$\overline{\rm M}$-conversion
would violate additive lepton family
number conservation and is discussed  in many speculative
theories (see Fig. \ref{theo_mmb}).It would be an analogy in the lepton
sector to ${\rm K}^0$-$\overline{\rm K^0}$ oscillations.

\begin{figure}[bt]
\unitlength 1.0cm
\begin{minipage}{3.5cm}
              \caption[]{\protect{\label{mmbarsetup}}
              Top view of the MACS
              (Muonium - Antimuonium - Conversion - Spec\-trometer)
              apparatus at PSI to search for
              ${\rm M}-\overline{\rm M}$ - con\-vers\-ion
              \protect{\cite{Abela_96}}.  }
\end{minipage}
\hfill
\unitlength 1.0cm
\begin{minipage}{8.0cm}
\begin{picture}(5.5,6.7)
\hspace{0.0 cm}
\mbox{
\epsfig{figure=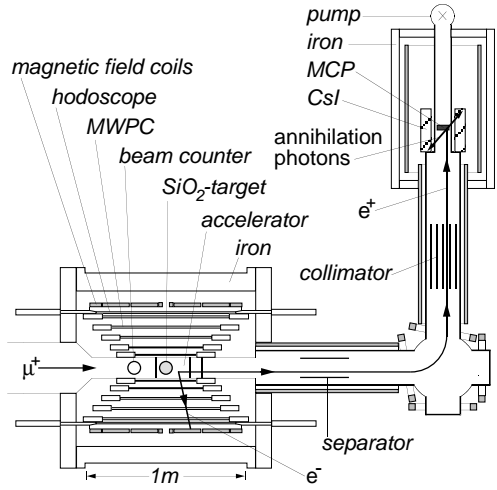,height=6.5cm,angle=90}
        }
\end{picture}
\end{minipage}
\end{figure}

The setup at PSI
(Fig. \ref{mmbarsetup}) \cite{Abela_96} is designed to employ the signature
developed
in a predecessor experiment at LAMPF, which requires the coincident identification
of both particles forming the antiatom in its decay
\cite{Matthias_91,Willmann_97}.
Muonium atoms in vacuum with thermal velocities, which are produced
from a SiO$_2$ powder target, are observed for
antimuonium decays. Energetic electrons from the decay of
the $\mu^-$ in the antiatom can be observed in a magnetic spectrometer
at 0.1~T magnetic field
consisting of five concentric multiwire proportional chambers
and a 64 fold segmented hodoscope.
The positron in the atomic shell of the antiatom
is left behind after the decay with 13.5~eV average kinetic energy \cite{chat92}.
It can be accelerated to 7~keV in a two stage electrostatic device and guided
in a magnetic transport system onto a position sensitive microchannel
plate detector (MCP). Annihilation radiation can be observed in a 12 fold
segmented pure CsI calorimeter around it.

\begin{figure}[bht]
\unitlength 1.0 cm
 \begin{minipage}{4.7cm}
 \begin{picture}(5.0,4.7)
  \centering
   \hspace{-0.7 cm}
   \epsfig{figure=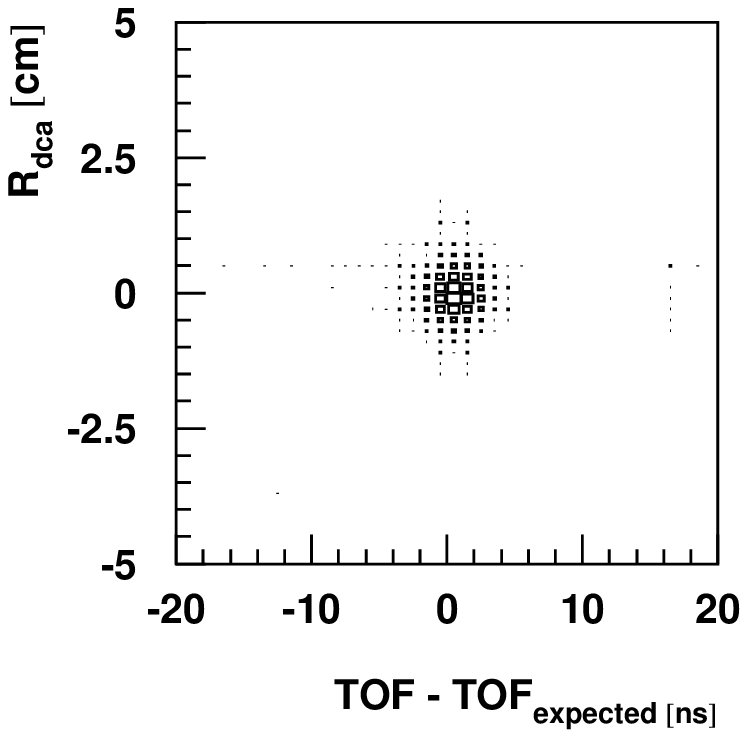,height=4.7cm}
 \end{picture}
 \end{minipage}
 \hspace{0.3cm}
 \begin{minipage}{4.7cm}
 \begin{picture}(5.0,4.7)
  \centering
   \epsfig{figure=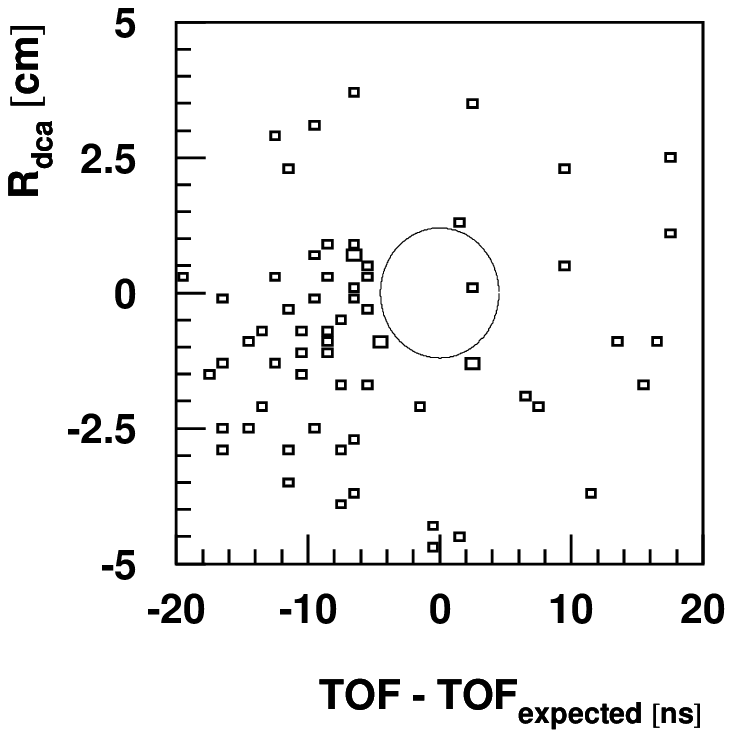,height=4.7cm}
 \end{picture}
 \end{minipage}
 \centering\caption[] {\protect{\label{res_mmb} }
         Time of flight (TOF) and vertex quality for a muonium measurement
        (left) and the same for all data of the final 4 month search for antimuonium
        (right). One event falls into the indicated 3 standard deviations area.
        }
\end{figure}

The relevant measurements were performed during in total 6 month
distributed over 4 years during which $5.7 \cdot 10^{10}$ muonium atoms
were in the interaction region.
One event
fell within a 99\% confidence interval of all relevant distributions
(Fig. \ref{res_mmb}).
The expected background due to accidental coincidences is 1.7(2) events.
Thus an upper limit on the conversion
probability
of ${\rm P_{M\overline{M}} \leq 8.2\cdot 10^{-11}}/ {\rm S}_{\rm B}$
(90\% C.L.) was found,
where ${\rm S}_{\rm B}$ accounts for the interaction type dependent
suppression of the conversion
in the magnetic field of the detector  due to the removal of degeneracy between
corresponding levels in M and ${\rm \overline{M}}$.
The reduction is strongest for (V$\pm$A)$\times$(V$\pm$A),
where ${\rm S}_{\rm B}$=0.35 \cite{hori96,wong95}.
This yields for the traditionally quoted upper limit
on the coupling constant in effective four fermion interaction
 ${\rm G_{M\overline{M}}} \leq 3.0\cdot 10^{-3} {\rm G}_{\rm F} (90\%C.L.)$
with ${\rm G}_{\rm F}$ the weak interaction Fermi constant.

This new result, which exceeds bounds from previous experiments
\cite{Matthias_91,Gordeev_94}
by a factor of 2500 and the one from an early stage of the experiment
\cite{Abela_96}
by 35,
has some impact on speculative models.
A certain  ${\rm Z}_8$ model is ruled out with more than 4 generations of particles
where masses could be generated radiatively with heavy lepton seeding
\cite{Wong_94}.

A new lower limit of $m_{X^{\pm \pm}}$ $ \geq $ 2.6~TeV/c$^2$ $*g_{3l}$ (95\% C.L.)
on the masses of flavour diagonal bileptonic gauge bosons in GUT models
is extracted which lies well
beyond the value derived from direct searches, measurements of the muon
magnetic anomaly or high energy Bhabha scattering
\cite{Sasaki_94,Cuyp_96}. Here  $g_{3l}$ is of order 1 and depends on the details of the
underlying symmetry. For 331 models this translates into
$m_{X^{\pm \pm}}$ $ \geq $ 850~GeV/c$^2$ which excludes their minimal Higgs
version in which an upper bound of 600~GeV/c$^2$ has been extracted from an analysis of
electroweak parameters \cite{fram97,fram97a}. The 331 models
may still be viable in some extended form involving a Higgs octet \cite{fram98}.
In the framework of R-parity violating supersymmetry \cite{moha92,Halp_93}
the bound on the coupling parameters could be lowered by a factor of 15
to $\mid \lambda_{132}\lambda_{231}^*\mid \leq 3 * 10^{-4}$ for assumed
superpartner masses of 100 GeV/c$^2$.
Further the achieved level of sensitivity allows to narrow slightly the
interval of allowed
heavy muon neutrino masses
in minimal left-right symmetry \cite{Herczeg_92} (where a lower bound on
${\rm G_{M\overline{M}}}$ exists, if muon neutrinos are heavier than 35 keV)
to $\approx$ 40~keV/c$^2$ up to the present
experimental bound at 170~keV/c$^2$.

In minimal left right symmetric models, in which ${\rm {M\overline{M}}}$ conversion is allowed,
the process is intimately connected to
the lepton family number violating muon decay
$\mu^+ \rightarrow e^+ + \nu_{\mu} + \overline{\nu}_e$.
With the limit achieved in this experiment
this decay is not an option
for explaining the excess neutrino counts in the LSND neutrino
experiment at Los Alamos  \cite{herc97,LSND_96}.

The consequences for atomic physics of muonium are such that the expected level
splitting in the ground state due to ${{\rm M} - \overline{{\rm M}}}$
interaction is below $1.5~{\rm Hz} / \sqrt{S_B}$ reassuring  the validity of
 fundamental constants determined in muonium spectroscopy.

A future  ${{\rm M} - \overline{{\rm M}}}$  experiment could take particularly
advantage of high intense pulsed beams.
In contrast to other LNV muon decays, the conversion through its nature
as particle - antiparticle oscillation,
has a time evolution in which the probability for finding
$\overline{{\rm M}}$ in the ensemble remaining after muon decay
increases quadratically in time, giving the signal an advantage growing in time
over major exponentially decaying background \cite{Willmann_97}.

\section{Long term future possibilities}

It appears that the availability of particles limits the ability to find very rare processes
or to impose significantly improved limits in continuation of the search program of dedicated experiments.
Therefore any measure to boost the respective particle fluxes is a very important
step forward.
The $\pi-\mu$ converter at PSI or the dedicated tailored muon production of the
planned MECO experiment at BNL are examples of
novel attempts to overcome this problem.
In principle, we need significantly more intense accelerators, such
as they are presently discussed at various places.
In the intermediate future the Japanese Hadron Facility (JHF) or a possible
European Spallation Source (ESS) are important options.
Also the discussed Oak Ridge neutron spallation source
could in principle accommodate intense muon beams.
The most promising facility would be, however, a muon collider \cite{Palmer_98},
the front end of which could provide
muon rates 5-6 orders of
magnitude higher than present beams (see Table \ref{muon_fluxes}).

%
\begin{table}[bht]
 \caption[]{\protect{\label{muon_fluxes}} Muon fluxes of
 some existing and future facilities, Rutherford Appleton Lab\-orat\-ory (RAL),
 Japanese Hadron Facility (JHF), European Spallation Source (ESS), Muon collider (MC). }
\begin{tabular}{|c|cccccc|}
\hline
                    &RAL($\mu^+$)    &PSI($\mu^+$)   & PSI($\mu^-$)  &JHF($\mu^+$)$^\dag$
                    &ESS($\mu^+$)     &MC ($\mu^+$, $\mu^-$)\\
 \hline
 Intensity ($\mu$/s)& $3\times 10^6$ &$3\times 10^8$ &$1\times 10^8$ &$4.5\times 10^7$
                     &$4.5\times 10^7$ &$7.5\times 10^{13}$ \\
 Momentum bite   &&&&&&\\
 \hspace*{4mm} $\Delta$ pm/p[\%] & 10& 10            & 10            & 10
  & 10              & 5-10     \\
  Spot size     &&&&&&\\
 (cm $\times$ cm)         & 1.2$\times$2.0 &3.3$\times$2.0  &3.3$\times$2.0  &
 1.5$\times$2.0 &1.5$\times$2.0 & few$\times$few \\
 Pulse structure    & 82 ns & 50 MHz    & 50 MHz    & 300 ns & 300 ns & 50 ps\\
                    & 50 Hz & continuous & continuous&  50 Hz &  50 Hz & 15 Hz\\
\hline
  \end{tabular}
  {\footnotesize
  $^\dag$ Recent studies indicate that the $10^{11}$ particles/s region might be reachable
  \cite{Kuno_98}.}
\end{table}

It was noted already in the early 60ies that, e.g. the process
$e^-e^- \rightarrow \mu^- \mu^-$ is closely related
to muonium-antimuonium conversion \cite{Glashow_61}.
Indeed such scattering experiments were carried out at the Princeton-Stanford
storage rings at Stanford yielding the at the time best limit on the coupling
 constant  $G_{{\rm M}\overline{\rm M}}$ \cite{Barber_69}. Today,
similar proposals have been made for scattering of high energy
$e^-$ on $e^-$,
$e^-$ on $e^+$,
$\mu^-$ on  $\mu^-$  and
$\mu^-$ on  $\mu^+$ \cite{Fram_92,Hou_96a,Raidal_98}. They were mainly discussed in
connection with bileponic gauge bosons. Even a lower limit for
the cross section of the  process $e^- e^- \rightarrow \mu^- \mu^-$
was found, provided
the sum of the light  neutrino masses exceeds $\approx 90$ eV \cite{Raidal_98}.
  Pronounced resonances have been predicted particularly for such experiments
at the Next Linear Collider or the high energy end of a muon collider. \\

Although lepton flavour conservation remains a mystery and searches
for its violation were not blessed with a successful observation yet,
both the theoretical  and experimental work in this connection have led
to a deeper understanding of
particle interactions. One  particular value of the
experiments are their continuos contributions towards
guiding theoretical developments by excluding various speculative models.

\acknowledgments{ It is a pleasure to thank the organizers of the
                  first tropical workshop for creating the atmosphere for a wonderful and
                  stimulating conference and for their great hospitality and support.
                  The author is grateful to M. Cooper, W. Molzon, A.v.d. Schaaf, M. Zeller
                  for discussions and updates, respectively latest results from their
                  experiments.}

\end{document}